\documentstyle[twocolumn,prl,aps,amssymb,epsfig]{revtex}
\begin{document}
\twocolumn[\hsize\textwidth\columnwidth\hsize\csname@twocolumnfalse\endcsname
\draft

\title{Morphological Symmetry Breaking During Epitaxial Growth \\at
Grazing Incidence}

\author{Jianxin Zhong,$^{1,2,*}$ Enge Wang,$^{3}$ Qian Niu,$^{4,2}$
and Zhenyu Zhang$^{2,1}$}

\address{ $^1$Department of Physics, University of Tennessee,
Knoxville, Tennessee 37996\\ $^2$Solid State Division, Oak Ridge
National Laboratory, Oak Ridge, Tennessee 37831\\ 
$^3$Institute of Physics \& Center of Condensed Matter Physics, 
Chinese Academy of Sciences, \\P. O. Box 603, Beijing 100080, P. R. China\\
$^4$Department of Physics, University of Texas at Austin, Austin, Texas 78712
}


\maketitle

\begin{abstract}%

It is shown that, in submonolayer growth at off-normal incidence, even
much less than one percent of transfer from the condensation energy of
the deposited atoms into adatom motion is sufficient to induce a net
adatom current from the illuminated edge of a two-dimensional island
to the other edges, thereby breaking the island symmetry.  Such a
symmetry breaking phenomenon is most pronounced for deposition at
grazing incidence.  Comparison between our theoretical predictions and
existing experimental results confirms the general validity of the
model.

\end{abstract}


]
\narrowtext

Because of their application potentials in future electronic devices,
a great deal of effort is being devoted to developing novel methods
for fabrication of organized low-dimensional structures, such as
ordered arrays of quantum wires or quantum dots\cite{HOM}. Epitaxial
growth and its inverse process, atom removal by sputtering or etching,
are two of the most promising approaches for mass production of
controlled nanostructures on various substrates. Typically, ordered
structures are obtained in heteroepitaxial systems, and the long-range
elastic field associated with the lattice mismatch between the two
systems plays an essential role in leading to self-organized growth
\cite{Tersoff}.  However, in recent studies of Cu(100) homoepitaxy, an
intriguing phenomenon about island formation and ordering has been
discovered, i.e., the average island symmetry varies with the incident
angle of deposition, leading to the formation of elongated islands, or
ripples, of twofold symmetry \cite{DJP}.  In contrast, earlier studies
of Cu(100) homoepitaxial growth at normal incidence resulted in only
square-shaped islands of fourfold symmetry \cite{DWZ,JBRP}.  Such a
symmetry breaking phenomenon is already present even in the
submonolayer growth regime, and is more dramatic at grazing incidence
\cite{DJP}.  For atom removal by sputtering under ion bombardment at
off-normal incidence, formation of coherent ripples has also been
observed on different substrates \cite{CB,CMK,RCBV}.

In earlier attempts to understand these symmetry breaking phenomena,
some qualitative suggestions have been proposed, all of which relying
on atoms climbing down from steps as an essential atomic process
\cite{DJP,RCBV}.  In this Letter, through a detailed study of a
simple model, we offer an alternative and quantitative interpretation
of the widely observed incidence geometry induced symmetry
breaking. Our model is based on an experimentally widely-invoked
concept, namely, transient mobility of deposited atoms on various
surfaces \cite{EJ,Tringides,ZMZW}. Through studying shape
evolution of monolayer-high islands on an fcc(100) surface, we
demonstrate that even much less than one percent of transfer of the
condensation energy from the deposited atoms into adatom motion is
sufficient to induce a net adatom current transferring adatoms from
the illuminated island edge to neighboring edges. As a consequence,
the symmetry of the growing islands changes from initially square
shape to rectangular shape elongated perpendicularly to the incident
direction.  Such a symmetry breaking phenomenon is most pronounced for
deposition at grazing incidence. A comparison between theory and
experiment confirms the general validity of the model. We also make
several specific predictions that can be tested in future experiments.

We use Fig.~1 to schematically show the growth process of a
monolayer-high island on an fcc(100) surface during deposition at
off-normal incidence. We denote the island width and length by
$W=m_xa$ and $L=m_ya$, respectively, where $a$ is the surface lattice
constant. We name the four island edges as $x^+$, $x^{-}$, $y^+$, and
$y^{-}$ edges. Atoms are deposited onto the surface at an angle
$\alpha$ with respect to the surface normal, with a deposition flux
$F$.  We consider the three most important kinetic processes for
adatom motion, i.e., the surface diffusion, the island edge diffusion,
and the island corner crossing, with rates $q_i=\nu_i\exp{(-V_i/kT)}$,
where $V_i$ and $\nu_i$ ($i=s,e,c$) are the corresponding barriers and
attempt frequencies. In general, one has $V_c=V_e+\Delta V$, with
$\Delta V >0$ because an adatom has to lower its coordination in
crossing an island corner \cite{ZL,Kellogg2}.

In the case of deposition at off-normal incidence, the effective flux
for depositing atoms on the surface is $F\cos\alpha$. The adatoms
diffuse on the surface and attach to an existing island with a flux
$f_s={F\cos{\alpha}\over N}$, where $N$ is the island
density. Following classical nucleation theory \cite{V}, we have
\begin{equation}
f_s= ({\nu_s a^2F^2\cos^2\alpha\over 3\theta})^{1\over 3}
\exp{(-{V_s\over 3kT})},
\end{equation}
where $\theta$ is the coverage.  Because of the isotropic nature of
adatom diffusion on an fcc(100) surface, the fluxes for adatom
attachment at the $x$ ($x^+$ or $x^-$) and $y$ ($y^+$ or $y^-$) island
edges are given by $f_x={m_x\over 2(m_x+m_y)}f_s$ and $f_y={m_y\over
2(m_x+m_y)}f_s$, respectively.  A remarkable feature for the
off-normal deposition is that it directly deposits atoms on the island
edge facing the incident beam (the illuminated edge), with a flux
$f_d=Fm_ya^2\sin{\alpha}$, valid in the low coverage limit.

\begin{figure}
\hspace{0cm}\rotatebox{-90}{ \resizebox{2.5in}{!}
{\includegraphics{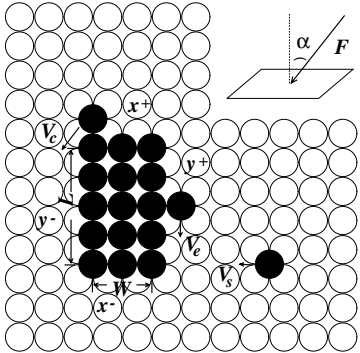}}} 
\\
\vspace{-1.5cm}
\caption{Schematic illustration of island growth on an fcc(100) surface
at off-normal incidence (see text for notations).}
\end{figure}

It is easy to conjecture on the island shape evolution if without the
contribution of the condensation energy.  Because of the existence of
the direct landing flux $f_d$, the illuminated edge (the $y^+$ edge in
Fig.~1) has more arriving atoms than the other three edges. In the low
temperature regime where island-corner crossing is not frequent enough
to establish equilibrium adatom distribution along the four edges, the
islands would elongate parallel to the incident direction.  In
contrast, at higher temperatures, frequent edge diffusion and corner
crossing can take place, leading to adatom transfer from one edge to the
other edges. This yields a uniform adatom distribution on the four
island edges. Therefore, without the contribution of condensation
energy, one would only find compact islands of either square shape or
rectangular shape elongated parallel to the incident direction.

Now, let us consider the effect of the condensation energy transfer to
adatom diffusion.  As stated above, the illuminated island edge has
more landing atoms than the other three edges.  The existence of the
transient mobility \cite{EJ,Tringides,ZMZW} implies that
those adatoms directly deposited at the illuminated edge will have
higher mobility to cross the two island corners bounding this edge.
Such anisotropic island corner crossing induces a net current
transferring adatoms from the illuminated edge to its neighboring
edges, and may therefore change the symmetry of the initially
square-shaped islands. As a result, the islands can elongate along the
direction perpendicular to the incident direction, which is just
the opposite to the expectation given above without consideration of
condensation energy transfer.

To quantify the above picture, next we derive a set of equations
describing the island evolution in the presence of condensation energy
transfer.  In a unit time, the total condensation energy given up by
the atoms directly deposited at the illuminated edge amounts to
$f_dU_0$, where $U_0$ is the condensation energy for each atom. During
the same time, the number of atoms landing on the illuminated edge is
$f_d+f_y$. Because only a portion of the condensation energy is
transferred into diffusional motion, we write the average energy gain
of an adatom on the illuminated edge as $\Delta E=\beta
f_dU_0/(f_d+f_y)$.  Here $\beta$ is a parameter describing the
transfer efficiency from the condensation energy of the incident atom
to its diffusional motion, with $\beta=1$ corresponding to complete
transfer and $\beta=0$ to zero transfer.  It should be noted that,
besides its being finite based on the experimental evidences for
transient mobility \cite{EJ,Tringides,ZMZW}, very little
is known about this parameter $\beta$. Nevertheless, it is known to
possess the following qualitative features, based on simple physical
considerations.  First, $\beta$ is weakly temperature-dependent for a
given system, because the condensation energy is much larger than the
thermal energy at typical growth temperatures.  Second, $\beta$ is
also weakly dependent on the incident angle, because the primary
angular dependence of the problem has already been incorporated into
the flux expression $f_d$.  Thirdly, $\beta$ is expected to be
strongly system-dependent, larger for a larger mass mismatch between
the incident atom and the substrate atom.  This last point is
transparent within the simple picture of elastic collision between the
incident atom and a surface atom, and is still qualitatively correct
even if the frictional forces due to vibrational and/or electronic
damping are included.

With the above considerations, here we can treat $\beta$ as a fitting
parameter, and approximate the higher rate of an adatom on the
illuminated island edge by $q_h=\nu_c\exp{(-{V_c\over kT+\Delta E})}$
when crossing an island corner to a neighboring edge.  The average
frequency for an adatom on an island edge of length $m$ to cross the
island corner with rate $q_c$ can be well approximated \cite{ZZZL} by
${1\over m}q_c$ when $V_c>V_e$.  Assume that the numbers of adatoms on
the $x^+$, $x^-$, $y^+$, and $y^-$ edges at time $t$ are $n_x^+$,
$n_x^-$, $n_y^+$, and $n_y^-$, respectively. Because of the higher
island corner crossing rate for adatoms on the illuminated island
edge, there must exist a net adatom current, $j_1$, which transfers
adatoms from the $y^+$ edge to its neighboring $x^+$ or $x^-$ edge. As
a consequence, one has $n_x^+>n_y^-$ and $n_x^->n_y^-$, which further
induce another net adatom current, $j_2$, carrying adatoms from the
$x^+$ and $x^-$ edges to the $y^-$ edge.  All together, the growth of
the island is described by
\[
{dn_y^+\over dt}=f_y+f_d-2j_1\]
\begin{equation}
{dn_y^-\over dt}=f_y+2j_2
\end{equation}
\[
{dn_x^+\over dt}={dn_x^-\over dt}=f_x+j_1-j_2, 
\]
where 
\begin{equation}
j_1={n_y^+\over m_y}q_h-
    {n_x^+\over m_x}q_c, {~~}
j_2=({n_x^+\over m_x}-{n_y^-\over m_y})q_c.
\end{equation}
Variations of the island width, $\Delta m_xa$, and island length,
$\Delta m_ya$, are given by $\Delta m_x={n_y^+ +n_y^-\over m_y}$ and
$\Delta m_y={n_x^+ + n_x^-\over m_x}$.  The ratio
$r\equiv {\Delta L\over \Delta W}=\Delta m_y/\Delta m_x$ predicts 
the evolution of the island
shape. A square-shaped island will  remain a square
if $r=1$ (stable growth), and will elongate perpendicularly to the
incident direction if $r>1$ or parallel to the incident direction if
$r<1$ (unstable growth).

\vspace{-1.5cm}
\begin{figure}
\hspace{0cm} ~~~~~~~~\rotatebox{-90}{\resizebox{2.5in}{!}
{\includegraphics{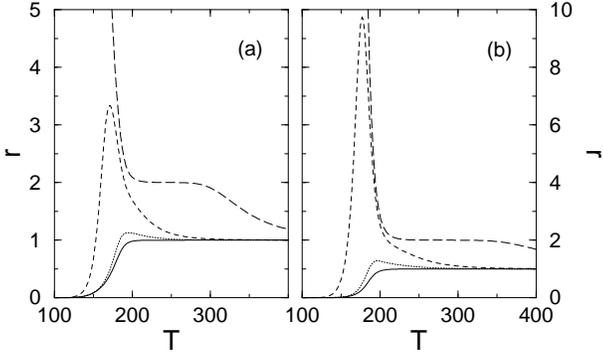}}} 
\\
\vspace{-0.5cm}
\caption{Temperature dependence of island growth on an fcc(100)
surface with different contributions of condensation energy. The
initial island size is $m_x^0=m_y^0=2$ in (a) and $m_x^0=2, m_y^0=8$
in (b). Atoms are deposited at a grazing angle of $\alpha =85^\circ$.
Solid, dotted, dashed, and long dashed lines correspond to $\beta=0,
0.0001, 0.0006$, and $0.05$, respectively.  Curves for $\beta=0.05$
are truncated for $r\geq 5$ in (a) and $r\geq 10$ in (b).}
\end{figure}

Because $\Delta E$ reaches its maximum when $\alpha$ approaches
$90^\circ$, the effect of the condensation energy is most pronounced
in the case of grazing incidence.  In the following, we first focus
our attention on the growth at grazing incidence.  Typical behaviors
of $r$ versus $T$ for different $\beta$ are shown in Fig.~2, where the
growth parameters are $F=0.1{\rm MLs}^{-1}$, $\nu_s=\nu_c=10^{12}{\rm
s}^{-1}$, $V_s=0.35$eV, $V_e=0.1$eV, $\Delta V=0.3$eV, $U_0=5$eV, and
$t=0.1$s.  Fig.~2(a) is for an initial square island of size
$m_x^0=m_y^0=2$. Fig.~2(b) is for an intermediate rectangular island
with $m_x^0=2$ and $m_y^0=8$.  As expected, without contribution of
the condensation energy, i.e., $\beta=0$, the ratio $r$ remains a
constant $r=1$ when $T>>T_c$ and $r<1$ when $T<T_c$, where $T_c\approx
190$K is the freezing temperature for island corner crossing defined
by $tq_c(T_c)\sim 1$.  For a non-zero $\beta$, we find three different
$r-T$ behaviors upon changing the energy transfer coefficient $\beta$.
In the case of a very small $\beta$, e.g., $\beta =10^{-4}$ in Fig.~2,
$r$ increases with decreasing temperature until it reaches
$T_c$. Further decreasing $T$ rapidly turns $r$ to zero. A notable
feature of $r$ in the temperature regime $T>T_c$ is $1<r<2$.  For
larger $\beta$, e.g., $\beta =6\times 10^{-4}$ in Fig.~2, when $T$
approaches $T_c$, $r$ increases from $r=1$ at high temperatures to a
ratio $r=2$. After that, a highly anisotropic growth with $r>2$
appears until the temperature reaches another critical value
$T_a\approx 170$K.  When $T$ further goes down, $r$ drops to zero. We
identify $T_a$ to be the freezing temperature for condensation energy
assisted island corner crossing defined by $tq_h(T_a)\sim 1$. In the
case of a much larger coefficient $\beta$, e.g., $\beta =0.05$, a
steady growth regime characterized by $r\equiv 2$ exists at
moderate temperatures, and the
strongly anisotropic growth mode $r>2$ remains in the whole
temperature range of $T<T_c$.  We note that the
steady growth regime exists for all $\beta >0.05$.  Figure 2 also shows
that the value of $r$ is larger for a larger value of $m_y^0/m_x^0$
(except for the steady state regime where $r\equiv 2$), indicating
that the elongation instability is more pronounced for already
developed rectangular islands.

\vspace{0cm}
\begin{figure}
\hspace{.5cm} \rotatebox{-90}{ \resizebox{2.2in}{!}
{\includegraphics{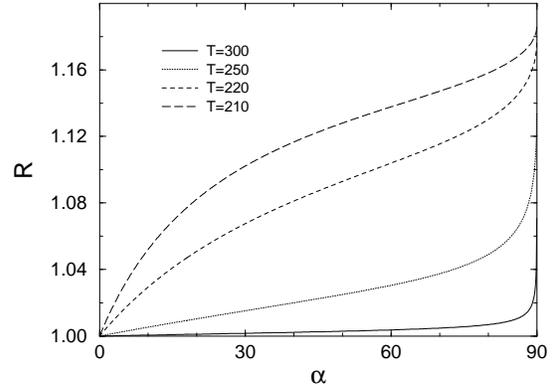}}} 
\\
\vspace{-0.5cm}
\caption{Island growth on a Cu/Cu(001) surface at off-normal incidence
and four different temperatures.}
\end{figure}

The observed different behaviors of $r$ for different $\beta$ can be
rationalized based on the competition of the three rate processes:
surface diffusion, thermally activated island corner crossing (without
the assistance of the condensation energy), and condensation energy
assisted island corner crossing.  In the temperature regime $T>T_c$,
frequent corner crossing with $q_c>>1$ and $q_h>q_c$ induces a net
current $j_1$ transferring adatoms from the illuminated edge to its
neighboring edges. On the other hand, $q_c>>1$ also results in a
uniform adatom distribution on the $x^+$, $x^-$, and $y^-$ edges, with
$j_2=0$. From Eqs. (2)-(3), we have $r=2/[1+(1+2m/n)\xi]$, where
$\xi=(f_y+f_d-2j_1)/(2f_x+f_y+j_1)$.  Because typically $V_s\leq V_c$,
surface diffusion is activated when $q_c>>1$, leading to $f_s>>f_d$.
Moreover, for a small $\beta$, the net current is small with
$j_1<(f_d+f_s)/2$.  Therefore, we have $1<r<2$ for $T>T_c$, as shown in
Fig.~2 for $\beta =10^{-4}$.  By increasing $\beta$, $j_1$ reaches the
maximum current $j_1=(f_s+f_d)/2$ upon decreasing $T$ toward $T_c$. In
this case, we find $r=2$, as shown in Fig.~2 for $\beta =6\times
10^{-4}$ and $\beta =0.05$. In the temperature regime $T<T_c$ for
small $\beta$, corner crossing and surface diffusion are frozen with
$q_c<<1$, $q_h<<1$ and $q_s<<1$. The landing atoms from the direct
flux $f_d$ accumulate on the illuminated edge, leading to $0\approx
r<1$. For $T_a<T<T_c$, we have $q_c<<1$ and $q_s<<1$ but
$q_h>>1$. Such a regime of strongly anisotropic island corner crossing
leads to $j_1=(j_s+j_d)/2$, which in turn results in strongly
anisotropic growth with $r>2$, as shown in Fig.~2 for $\beta =6\times
10^{-4}$. Finally, for even larger $\beta$, we have $q_h>>1$ for any
temperature below $T_c$, leading to $r>2$ for $T<T_c$ (see Fig.~2 for
$\beta=0.05$). The above discussions are valid as long as the islands
are compact \cite{ZZZL}.

Finally, through studying Cu/Cu(001) growth, we examine how the aspect
ratio $r$ depends on the incident angle. Fig.~3 is obtained for
various temperatures with the same flux $F=0.0042$MLs$^{-1}$ as used
in the experiments \cite{DJP}. The barriers are $V_s=0.505$ eV,
$V_e=0.265$ eV, and $\Delta V=0.29$ eV, as suggested by
experiments\cite{DWZ} and embedded-atom model calculations\cite{ZZZL}.
The condensation energy \cite{EJ} is taken as $U_0=3$eV and a
reasonably small energy transfer coefficient, $\beta=8\times 10^{-5}$,
is used.  Figure 3 indicates that, at room temperature, the elongation
is not easily measurable for smaller $\alpha$.  Significant change of
the island shape occurs at larger grazing angles ($\alpha >80^\circ$).
This explains why one usually observes square-shaped islands during
deposition at normal incidence or small-angle incidence \cite{DWZ,JBRP} but
elongated islands at grazing incidence \cite{DJP}. Furthermore, after
integration of Eq.~(2) to a coverage $\theta=0.5$, we found that the
aspect ratio has a value $r\equiv {L\over W}=1.05$ for $T=250$ K at
$\alpha=80^\circ$, as found in the experiments \cite{DJP}.  Moreover,
Fig.~3 shows that $r$ can be greatly enhanced by slightly decreasing
the growth temperature, a prediction to be confirmed in future 
experiments.  We also note that if $\beta$ increases weakly with
$\alpha$, the crossover shown in Fig. 3 will be even sharper.

In summary, through studying monolayer-high island shape evolution on
an fcc(100) surface, we have shown that the condensation energy of
deposited adatoms can play an important rule in controlling the island
shape during epitaxial growth.  In the case of deposition at
off-normal incidence, the component of the deposition flux parallel to
the surface provides additional atoms on the illuminated island edge
and thus more condensation energy. This leads to an enhancement of the
mobility of the adatoms on the illuminated edge, and results in island
elongation perpendicular to the incident direction.  Such an island
symmetry breaking phenomenon is most pronounced at grazing
incidence. A comparison between the theoretical predictions and the
experimental findings in Cu/Cu(001) growth confirms the general
validity of the model.  We have also found strong temperature
dependence of the aspect ratio, and the existence of a well-defined
incident angle above which the elongation instability is most
pronounced. These latter predictions are to be verified in future
experiments.

\vspace{0.5cm} We thank Dr. W.F. Egelhoff, Jr. for a careful reading of
the manuscript, and for his helpful comments and suggestions.  This
research was supported by the US National Science Foundation under
Grant No. DMR-9702938, by Oak Ridge National Laboratory, managed by
Lockheed Martin Energy Research Corp. for the Department of Energy
under contract No. DE-AC05-96OR22464, and by the Natural Science
Foundation of China under Grant Nos. 19810760328 and 19928409.

\vspace{0.5cm}
\noindent
$^*$ On leave from the Department of Physics, Xiangtan University, 
Hunan 411105, China.

\end{document}